
\magnification 1200
\def\sqr#1#2{{\vcenter{\hrule height.#2pt width 6pt
    \hbox{\vrule width.#2pt height#1pt \kern#1pt
    \vrule width.#2pt}
   \hrule height.#2pt width 6pt}}}

\def\section#1#2{\medskip\noindent{\bf#1\quad
  #2}\par\smallskip}
\def\Idoubled#1{{\rm I\kern-.22em #1}}
\def\b{{\beta}}
\def\d{{\delta}}
\def\D{{\Delta}}
\def\G{{\cal G}}
\def\w{\wedge}
\def\half{{1\over2}}
\def\a{\alpha}
\def\th{\theta}
\def\tA{{\tilde A}}

\def\fd{{\det{}}}

\def\b{{\beta}}
\def\dmu{{\cal D}\mu}
\def\ps{\psi_{(2)}}
\def\ph{\psi_{(n-1)}}
\def\bntwo{B_{(n-2)}}
\def\aone{A_{(1)}}
\def\at#1{{a^T_{(#1)}}}
\def\azero{\alpha_{(0)}}
\def\bl#1{{b^L_{(#1)}}}
\def\betal{{\beta^L_{(2)}}}
\def\gauge{\displaystyle{{\dmu[\at{n-3}] \dmu [\azero] \over
V_G}}}
\def\dA{d_{\tA}}
\def\dltA{\d_\tA}
\def\dltth{{\d_\th}}

\baselineskip 15 pt
\raggedbottom
\magnification 1200
\def\sqr#1#2{{\vcenter{\hrule height.#2pt width 6pt
    \hbox{\vrule width.#2pt height#1pt \kern#1pt
    \vrule width.#2pt}
   \hrule height.#2pt width 6pt}}}

\def\section#1#2{\medskip\noindent{\bf#1\quad
  #2}\par\smallskip}
\def\Idoubled#1{{\rm I\kern-.22em #1}}
\def\b{{\beta}}
\def\d{{\delta}}
\def\D{{\Delta}}
\def\G{{\cal G}}
\def\w{\wedge}
\def\half{{1\over2}}
\def\a{\alpha}
\def\th{\theta}
\def\tA{{\tilde A}}

\def\fd{{\det{}}}

\def\b{{\beta}}
\def\dmu{{\cal D}\mu}
\def\bntwo{B_{(n-2)}}
\def\aone{A_{(1)}}
\def\at#1{{a^T_{(#1)}}}
\def\azero{\alpha_{(0)}}
\def\bl#1{{b^L_{(#1)}}}
\def\betal{{\beta^L_{(2)}}}
\def\gauge{\displaystyle{{\dmu[\at{n-3}] \dmu [\azero] \over
V_G}}}
\centerline{{\bf THE PARTITION FUNCTION FOR}}
\centerline{{\bf TOPOLOGICAL FIELD THEORIES}}
\vskip 20 pt
\centerline{{\it by}}
\vskip 20 pt
\centerline{{\bf J. Gegenberg${}^\dagger$
      \  G. Kunstatter${}^\sharp$}}
\vskip 20 pt
\centerline{{\it${}^\dagger$Dept. of Mathematics and Statistics,
University of
New
Brunswick}}
\centerline{{\it Fredericton, New Brunswick}}
\centerline{{\it Canada E3B 5A3}}
\vskip 0.17 true cm
\centerline{{\it ${}^\sharp$ Winnipeg Institute for Theoretical
Physics
and Physics Dept.}}
\centerline{{\it University of
Winnipeg}}
\centerline{{\it Winnipeg, Manitoba}}
\centerline{{\it Canada R3B 2E9}}
\vskip 30 pt
\noindent{\bf ABSTRACT}
\smallskip
\noindent
We use a Hodge decomposition and its generalization
to non-abelian flat vector bundles to calculate the partition
function for abelian and non-
abelian BF theories in $n$ dimensions.   This enables us to
provide a simple proof that the partition
function is related to the Ray-Singer
torsion defined on flat vector bundles for all odd-dimensional
manifolds, and is equal to unity for even dimensions.
\bigskip\centerline{June 1992, revised March 1993}\bigskip\noindent
WIN-92-03
\par\noindent
UNB Technical Report 92-01
\vfill
\eject
\section{1.}{INTRODUCTION}
Topological field theories, such as Einstein gravity in
2+1 spacetime dimensions or the so-called
BF theories[2, 11,12,13] are characterized by the
following properties:
First, the gauge group is sufficiently large so as to
eliminate all local dynamics.
Second, physical quantities are independent of the
choice of spacetime metric.  The net result is that the
construction of quantum versions of these theories proceeds much
like the construction of quantum {\it mechanical} systems[4,11].
\par
Topological field theories are of interest for at least two
important reasons. First they provide examples of quantum field
theories in which the non-perturbative sector can be examined
quantitatively. It is therefore hoped that these simpler theories
will yield insight into the non-perturbative quantization of more
complicated, and physically relevant theories.
An important example in this regard is quantum gravity.  A
complete quantum gravity theory in 3+1 spacetime dimensions does
not yet exist, despite considerable recent progress by Ashtekar
and his group (for a recent review, see Reference 3) in
the canonical quantization.  Einstein gravity in 2+1
spacetime dimensions, on the other hand, is a topological field
theory which is completely solvable, both
classically and quantum mechanically[4, 5,
6, 7, 8].  From the latter, some progress has been made in
solving important technical problems(e.g. the construction of
observables[9]) and in sorting out deep conceptual
issues(e.g., the nature of time[10]).
\par
Second, topological field theories have provided a new framework
for understanding important topological invariants.  To cite two
examples, we mention first the discovery by  Horowitz and
Srednicki[12] and by Blau and Thompson[2] that the linking number
of two surfaces can be represented by a correlation function
computed with respect to a quantized abelian BF theory; and
second,  Witten's representation of knot polynomials as
expectation values
of Wilson loop observables with respect to quantized non-abelian
Chern-Simons theories[14].
\par
The partition function is an important tool for topological
quantum field theory, as well as for traditional quantum field
theory.  In the latter it plays a key part in the determination
of
the perturbative structure, while in the former it yields
important topological invariants[2,13,14,15]. The most serious
impediment to the calculation of the partition function (and of
course other functional integrals) is the degeneracy of the
lagrangian density
due to the gauge invariance of the action, and in topological
field theories, as in traditional
quantum field theories, the technique employed is some variant of
the Faddeev-Popov {\it ansatz}[16].  Schwarz[13] used a method
(dubbed the ``method of resolvents") based on the latter to
compute the partition function for abelian BF theory,
in which the fields reside in a vector bundle with a given flat
connection, in terms of the Ray-Singer torsion associated with
that flat connection.  However, Horowitz and Srednicki[12]
computed correlation functions for the same class of theories as
Schwarz by simply parameterizing the fields by their Hodge
decomposition.  There was no need to gauge fix since the
Hodge decomposition {\it explicitly} parameterizes the gauge and
the physical degrees of freedom, and the Jacobian of the
transformation to the latter can be read off from the inner
product.
\par
In their review of topological field theories,
Blau and Thompson[2] note that the Fadeev-Popov {\it ansatz}
applied
to the computation of the partition function for the abelian
BF models amounts to factoring out and cancelling the integral
over the exact parts and functionally integrating over the
co-exact parts of the Hodge decomposition of the fields.  This
provides them with a preferred gauge choice, which they then use
to compute the partition function via the BRST procedure.
\par
The above discussion suggests the use of the Hodge
decomposition of the fields as a parameterization in which the
partition function for topological field theories can be directly
calculated without encountering the usual complications that
arise in the BRST procedure. The purpose of
this paper is to show that this suggestion can be ``rigorously"
implemented for abelian Chern-Simons
and for both abelian and non-abelian BF theories in arbitrary
dimensions. As expected, in the abelian case the technique
reproduces the results of Schwarz[13]. However, in the non-abelian
case, the method allows us to write down for the first time, to
our knowledge, the partition functions for non-abelian BF theory
in $n$-dimensions. In particular we are able to prove the
triviality of the partition function in even dimensions, and in odd
dimensions to relate the partition function to a sum over
Ray-Singer torsions
associated with the equivalence class of flat connections
parametrized by the moduli space of the theories.
\par
The paper is organized as follows:
Section 2. reviews the Hodge decomposition and evaluates the
partition function for abelian Chern-Simons and $n$-dimensional
BF theory.
Section 3. presents the generalization of these methods to
non-abelian BF theories. The proofs of some important identities
are relegated to an Appendix. Section 4. closes with a summary
and discussion.
\eject
\section{2.}{ABELIAN THEORIES}
\medskip
In this Section, it is shown that a Hodge decomposition allows
one to compute
functional integrals
for abelian BF and Chern-Simons models in a straightforward
manner.  This method is equivalent to Schwarz' method of
resolvents[2,13] for abelian BF models. Similar techniques
have been used by Horowitz and Srednicki[12] to compute linking
numbers.
\section{2.1.}{The Parametrization}
First, let us recall the Hodge decomposition of a differential
form[17].  Given a smooth $p$-form $\omega_{(p)}$ on a closed
compact orientable differentiable manifold $M$ of dimension
$n\geq p$, then
there exist smooth differential forms
$\a_{(p-1)},\b_{(p+1)},h_{(p)}$ such that
$$\omega_{(p)}=h_{(p)}+
d\a_{(p-1)}+\d\b_{(p+1)},\eqno(2.1)$$
where $d$ denotes exterior differentiation, $\delta\equiv (-
1)^{np+n+1}*d*$, with $*$ denoting the Hodge dual.  Note that
$**\omega_{(p)}=(-1)^{p(n-p)}\omega_{(p)}$.  The manifold
$M$ must be provided with a metric tensor in order to define
Hodge duality, co-derivatives and Laplacians[18].  In (2.1)
 $h_{(p)}$ is {\it harmonic}, i.e.,
$$dh_{(p)}=0=\d
h_{(p)},\eqno(2.2)$$
so that
$$\D_{(p)}h_{(p)}:=(d\d+\d d)h_{(p)}=0.$$
 The exterior derivative $d$ and the
corresponding co-derivative $\d$ are nilpotent:  $dd=0=\d\d$.
The terms $d\a_{(p-1)}$ and $\d\b_{(p+1)}$ are called the {\it
exact} and {\it co-exact} parts, respectively, by mathematicians;
or the   {\it
longitudinal} (or {\it pure gauge}) and {\it transverse} parts,
respectively, by physicists.  It is important that the harmonic,
longitudinal and transverse parts of the $\omega_{(p)}$ are
unique, providing that we exclude zero modes of the operators
$(\d d)$ and $(d\d)$ from the set of forms $\a_{(p-1)}$ and
$\b_{(p+1)}$, respectively. Moreover, we require that the forms
$\alpha_{(p-1)}$ and $\b_{(p+1)}$ are themselves uniquely
specified and that they can be explicitly represented. This in
turn can be achieved by requiring $\alpha_{(p-1)}$ and
$\b_{(p+1)}$  to be transverse and longitudinal, respectively.
That is, if we consider the Hodge decomposition of $\alpha_{(p-
1)}$, we find that the longitudinal part of
$\omega_{(p)}$, namely $d\a_{(p-1)}$ is equal to
$d\d\gamma_{(p)}$,
where $\d\gamma_{(p)}$ is the {\it unique} transverse part of
$\a_{(p-1)}$.  Hence we can parametrize $\a_{(p-1)}$ by its
unique  transverse part.  This parametrization can be made
explicit by expanding $\gamma_{(p)}$ in terms of the eigenmodes
(excluding zero modes) of the operator $d\d$. Similarly we can
parametrize $\beta_{(p+1)}$ by its unique longitudinal part.
\par
The harmonic part $h_{(p)}$ depends
only on the global structure, i.e. the topology, of $M$.  Indeed,
for
a given $p$, the space of harmonic $p$-forms is a vector space
isomorphic to the de Rham cohomology of $M$ [17,18].  \par In the
following discussion it will be necessary to define a functional
measure on the space of $p$-forms. This can be done by first
noting
the existence of a natural inner product between any two
$p$-forms,
$\omega_{(p)}$ and $\omega'_{(p)}$[12]:  $$
<\omega_{(p)}\vert\omega'_{(p)}>:=\int_M\omega_{(p)}
\wedge*\omega'_{(p)}.  \eqno(2.3)$$ This leads to a natural
functional measure $\dmu[\omega_{(p)}]$ on the space of
$p$-forms,
namely that which normalizes the Gaussian functional integral:
$$
\int \dmu[\omega_{(p)}] e^{i <\omega_{(p)}\vert
  \omega_{(p)}>}=1.  \eqno(2.4)$$ Note that both the inner
product
(2.3) and hence the functional measure in (2.4) depend explicitly
on
the metric on $M$. Even for compact closed $M$, the functional
measure (2.4) is formally infinite and requires regularization.
\par
One can use the invariance of (2.4) to determine how the
functional
measure transforms under the reparametrization (2.1). Using (2.1)
and
the orthogonality of the Hodge decomposition with respect to the
inner product (2.3) one can show that:
$$\eqalignno{<\omega_{(p)}\vert\omega_{(p)}>&=
    <h_{(p)}\vert h_{(p)}> + <d\a_{(p-1)}\vert d\a_{(p-1)}> +
 <\delta\b_{(p+1)}\vert\delta\b_{(p+1)}>\cr &= <h_{(p)}\vert
h_{(p)}>
 + <\a_{(p-1)}\vert(\delta d)\a_{(p-1)}> +<\b_{(p+1)}\vert
  (d\delta)\b_{(p+1)}>,&(2.5)\cr}$$ where we have used the
identities
[17]:  $$ <d\a\vert d\a>=
   <\a\vert(\delta d)\a>, \eqno(2.6a)$$ and $$ <\delta
\b\vert\delta\b>=<\b\vert (d\delta)\b>.  \eqno(2.6b)$$ Since, for
any
linear operator $O$, one has:  $$ \int \dmu[\omega_{(p)}] \exp
i<\omega_{(p)}\vert O  \omega_{(p)}>= \fd^{-\half}(O),
\eqno(2.7)$$
(2.4) and (2.5) imply that $$\dmu[\omega_{(p)}]=
\dmu[h_{(p)}]\dmu[\a_{(p-1)}]\dmu[\b_{(p+1)}] \fd^\half (\delta
d)_{(p-1)} \fd^\half (d\delta)_{(p+1)}.  \eqno(2.8)$$ Note that
uniqueness of the Hodge decomposition requires that the zero
modes of
the relevant operators are excluded from the parametrization so
that
the Jacobian of the transformation is non-singular (when
regulated
appropriately).  \medskip \section{2.2.}{Abelian Chern-Simons
Theory}
The classical action for an abelian Chern-Simons theory,
$$S=\int_M A\w
dA\,\,, \eqno(2.9)$$ is invariant (up to a total divergence)
under
the gauge transformation:  $$A\rightarrow A+d\Lambda.
\eqno(2.10)$$
We wish to compute the partition function for the theory:
$$Z:=\int
{\dmu[A]\over V_G} e^{i S[A]}\,\,.  \eqno(2.11)$$ In the above
$V_G$
denotes the volume of the group of gauge transformations in
(2.10),
which must be factored out of the partition function in order to
guarantee that the integration is performed only over physically
distinct gauge fields.  The normal method for handling this is to
add
a lagrange multiplier term enforcing a gauge fixing condition to
the
classical action, and then to use the Fadeev-Popov {\it ansatz},
or
its generalizations to add compensating ghost terms to the
functional
integral in order to guarantee that only physical modes
contribute[16].  Alternatively, as we shall now demonstrate for
the
simple case of abelian Chern-Simons theory, the same results can
be
obtained by using the Hodge decomposition to parametrize the
potential $A$ in terms of its gauge invariant, and gauge
dependent
parts,  so that  the volume of the group  of gauge
transformations
can be explicitly factored out, leaving a functional integral
over
gauge invariant modes only.  \par We now transform the
integration
variables:  $$A\rightarrow \a,\b, h,$$ where $\a,\b, h$
parameterize
respectively the exact, co-exact, and harmonic parts  of the
connection A.  Taking into account the Jacobian (2.5) we get:
$$Z=\int{\dmu[\a] \dmu[\b] \dmu[h]\over
V_G}\fd^{\half}\D_{(0)}\fd^{\half}(d\d)_{(2)}e^{iS}\,\,,
\eqno(2.12)$$ where we have used the fact that $\Delta_{(0)}=(\d
d)_{(0)}$.  In terms of this parametrization, it is clear that
the
volume of the gauge group is simply
 $$V_G=\int \dmu[\a],  \eqno(2.13)$$ while the classical action
 functional becomes, after integrating by  parts, using the
properties (2.2) of $h$ and the nilpotency of the exterior
derivative
operators, and dropping surface terms:  $$S=-<\b\vert (*\delta
d\delta) \b>\,\,.  \eqno(2.14)$$ Note that $S$ depends only the
co-exact (transverse) part of $A$.  Using (2.13) and integrating
over
$\beta$ yields:  $$Z=\int \dmu[h] \fd^{-\half}\left(*\d
d\d\right)_{(2)} \fd^{\half}\D_{(0)} \fd^{\half}(d\d)_{(2)}.
\eqno(2.15)$$ In the Appendix we prove that $$\fd(*\d d\d)_{(2)}=
\fd^\half((d\d d)(\d d\d))_{(2)} =\fd^{3\over2}(d\d)_{(2)}.
\eqno(A.1)$$ Using the identity $$\fd(\d
d)_{(p)}=\fd(d\d)_{(n-p)},
\eqno(2.16)$$ which is a consequence of Hodge duality, it follows
that $$Z=\int
\dmu[h]\fd^{-{3\over4}}\D^T_{(1)}\fd^{\half}\D_{(0)}
\fd^{\half}\D^T_{(1)}\,\,.  \eqno(2.17)$$ The operator
$\D^T_{(p)}$
is the transverse part of the laplacian acting on $p$-forms:
$$\D^T_{(p)}:=(\d d)_{(p)}.  \eqno(2.18)$$ Applying identity
(A.3)
proved in the Appendix, we have
$\fd\D^T_{(1)}=\fd\D_{(1)}/\fd\D_{(0)}$ and hence $$Z=\int\dmu[h]
\fd^{-{1\over4}}\D_{(1)}\fd^{3\over4}\D_{(0)}.  \eqno(2.19)$$ The
space of harmonic forms (of any order) consists of a finite set.
Hence, the integration over harmonic forms in the above is a
simple
sum.  We recognize the integrand in the above as the inverse of
the
square root of the Ray-Singer torsion of the de Rham cohomology
of
$M$ [2,13].
\eject
\section{2.3.}{Abelian BF Theories} The action for
abelian BF theories is[2] $$ S=\int \bntwo \wedge d\aone = -
<\bntwo\vert*d\aone>, \eqno(2.20)$$ which is invariant under the
gauge transformations:  $$\eqalignno{ \bntwo&\to\bntwo+d
a_{(n-3)},&(2.21a)\cr \aone&\to\aone+d\azero.&(2.21a) \cr}$$ The
partition function for the theory is therefore:  $$ Z= \int
\displaystyle{{\dmu[\bntwo]\dmu[\aone]\over V_G}}
 \exp i S[\bntwo,\aone], \eqno(2.22)$$ where the volume of the
gauge
group $V_G$ will be evaluated below.  \par We now do a Hodge
decomposition on the fields:  $$ \bntwo=h_{(n-2)}+ d\at{n-3}+\d
\bl{n-1}, \eqno(2.23)$$ $$ \aone= h_{(1)} + d\azero + \d \betal.
\eqno(2.24)$$ Note that we have added the superscripts $T$, $L$
to
the Hodge parameters to remind ourselves that they are
necessarily
transverse or longitudinal, respectively.  \par In terms of the
new
parameters, the action is, up to a sign:
$$\eqalignno{
S&=-<\bntwo\vert * d\aone>\cr
&=-<\d\bl{n-1}\vert *d\d\betal>\cr
&=-<\bl{n-1}\vert (*\d d\d) \betal>.&(2.25)\cr}$$ Taking the
Jacobian
into account, the partition function is $$\eqalignno{ Z&=\int
\dmu[h_{(n-2)}]\dmu[h_{(1)}] \gauge \dmu[\bl{n-1}]\dmu[\betal]\cr
&\quad\times \fd^\half(\d d)_{(n-3)} \fd^\half (d\d)_{(n-1)}
  \fd^\half\Delta_{(0)} \fd^\half(d\d)_{(2)} \quad
e^{iS}.&(2.26)\cr}$$
Integrating over $\bl{n-1)}$ yields the delta function:
$\delta(*\d
d\d \betal)$ so that, using the functional analogue of $$ \int dx
\,\delta(f(x)) ={1\over f'(0)},$$ the integral over $\betal$
yields:
$$\eqalignno{ Z&=\int \dmu[h_{(n-2)}]\dmu[h_{(1)}]
\gauge\fd^\half(\d
d)_{(n-3)} \fd^\half (d\d)_{(n-1)}\cr &\quad\times
\fd^\half\Delta_{(0)} \fd^\half(d\d)_{(2)}
  \fd^{-1}(*\d d\d)_{(2)}. &(2.27)\cr}$$ By using the identities
(A.1) and $$ \fd(d\d)_{(n-1)}=\fd (\delta
d)_{(n-2)}=\fd(d\d)_{(2)},
$$ which follow from (A.2), we get $$\eqalignno{ Z&=\int
\dmu[h_{(n-2)}]\dmu[h_{(1)}] \gauge \cr
 & \quad\times \fd^\half(\d d)_{(n-3)} \fd^{-\half} (\d
d)_{(n-2)}
\fd^\half(\d d)_{(0)} &(2.28).\cr}$$ \par The final step in the
calculation is to evaluate $V_G$. From Equations (2.21) it is
clear
that:  $$ V_G= V_{\bntwo}\times V_{\aone}, \eqno(2.29) $$ where,
as
in the case of  Chern-Simons theory:  $$
V_{\aone}=\int\dmu[\azero].
\eqno(2.30)$$ By analogy, one would be tempted to assume that:
$$
V_{\bntwo} = \int\dmu [\at{n-3}]. \eqno(2.31) $$ However, this is
not
consistent with the Faddeev-Popov ansatz for gauge fixing, which
in
turn is justified by a Hamiltonian analysis[16]. In particular,
(2.31) would result in the wrong mode count for BF theories and
also
for dynamical theories with reducible symmetries[19].  \par The
Faddeev-Popov ansatz can nonetheless be given a very simple
geometrical interpretation in terms of the considerations above,
and
the calculation of $V_{\bntwo}$ turns out to be remarkably
simple.
Gauge fixing via Faddeev-Popov is equivalent to considering
$V_{\bntwo}$ to be the volume of all possible (non-harmonic)
$(n-3)$
forms, $a_{(n-3)}$, modulo the group of transformations on
$a_{(n-3)}$ that leave $da_{(n-3)}$ invariant[16].  Since the
ghost-for-ghost prescription[23] treats these residual
transformations  in essentially the same way as the original
transformations (i.e. it adds gauge fixing terms, compensating
ghosts, etc.) it becomes apparent that $$
V_{\bntwo}=\displaystyle{{\int\dmu[a_{(n-3)}]\over
  V_{B_{(n-3)}}}}. \eqno(2.32) $$ This equation can be easily
iterated to yield\footnote{$^1$}{A similar result was obtained by
Obukhov[20] for antisymmetric tensor field theory.}:  $$
V_{B_{(n-2)}} = \prod^n_{i=3} \left[ \int
\dmu
 [a_{(n-i)}] \right]^{(-1)^{(i+1)}}. \eqno(2.33) $$ We now
express $$
a_{(n-i)} = \at{n-i} +d\at{n-i-1}, \eqno(2.34) $$ where we have
ignored harmonics. This is because they do not contribute
directly to
the gauge group and can be factored out of the integral over
$a_{(n-3)}$ without introducing any Jacobians.  The decomposition
(2.34) yields the following functional measure:  $$ \dmu
[a_{(n-i)}]
= \dmu [\at{n-i}] \dmu [\at{n-i-1}]
   \fd^\half (\d d)_{(n-i-1)}. \eqno(2.35) $$ Inserting (2.35)
into
(2.33) yields the following simple expression:  $$ V_{B_{(n-2)}}=
\int \dmu[\at{n-3}]
  \prod^{(n-1)}_{i=3} \fd^{\half(-1)^{i+1}}
   (\d d)_{(n-i-1)}, \eqno(2.36) $$ so that the partition
function
becomes:  $$ Z=
\int\dmu[h_{(n-2)}]\dmu[h_{(1)}]\fd^\half\Delta_{(0)}
\prod^{n}_{i=2}\fd^{\half(-1)^{i+1}}
   (\d d)_{(n-i)}. \eqno(2.37) $$
\par It is straightforward to
relate (2.37) to the expressions obtained previously by
Schwarz[13]
and by Blau and Thompson[2]. It is useful to write out the
product in
(2.37) explicitly:  $$\eqalignno{
Z&=\int\dmu[h_{(n-2)}]\dmu[h_{(1)}]\fd^\half\D_{(0)}
\fd^{-\half} (\d d)_{(n-2)} \fd ^\half (\d d)_{(n-3)}...\cr
&\qquad...\fd^{\half(-1)^{n-1}}(\d d)_{(2)}\fd^{\half(-1)^n}
  (\d d)_{(1)} \fd^{\half(-1)^{n+1}}\Delta_{(0)}.&(2.38)\cr}$$ In
even dimensions, $n=2m$, the identity (A.12) clearly guarantees
the
pairwise cancellation of all the determinants in the above
expression, so that the partition function is unity. In odd
dimensions, $n=2m+1$, the corresponding factors have exponents
with
the same sign, and do not cancel. Instead, one obtains:  $$
Z=\int\dmu[h_{(n-2)}]\dmu[h_{(1)}]
\fd^{\half(-1)^m} (\d d)_m\prod^{m-1}_{i=0} \fd^{(-1)^i}(\d d)_i
\,\,\, .\eqno(2.39) $$
One can now use (A.3) to write this in terms of
Laplacians:  $$ Z= \int\dmu[h_{(n-2)}]\dmu[h_{(1)}]\prod^m_{j=0}
\fd^{\nu_j}\Delta_{(j)},\eqno(2.40)
$$ where $\nu_j:= (-1)^j (m-j+\half)$.  The expression for the
partition function given by Blau and Thompson[2], for example, is
considerably more complicated:  $$
Z=\int\dmu[h_{(n-2)}]\dmu[h_{(1)}]\fd^{\tilde{\nu}_0}\Delta_{(1)}
\fd^{\tilde{\nu}_1} \Delta_{(0)} \prod^{2m-1}_{j=0}
\fd^{\tilde{\nu}_j} \Delta_{(2m-1-j)} \,\, , \eqno(2.41)$$ where
$\tilde{\nu}_j:= (-1)^{j+1}(2j+1)/4$.  However, by using Hodge
duality
(A.11), it is easy to show that (2.41) reduces to (2.40) given
above.  \eject
\section{3.}{Non-Abelian Theories} In this
section, we
will calculate the partition function for non-Abelian
BF
theories in n-dimensions, using techniques analoguous to those in
the
previous section. The action for these theories takes the
form[2]:
$$ S[A,B]=Tr\int_M B\w F(A),\eqno(3.1)$$ where $B$ is an
$(n-2)-$form
on $M$, taking its value in the Lie algebra ${\cal G}$ of a
semisimple
Lie group $G$ and $A$ is a $G$- connection 1-form field on $M$.
The
curvature of $A$, $F(A)$ is defined by $$ F(A):=dA+A\w A=dA+\half
[A,A].  \eqno(3.2)$$ \par The group $G$ generates the usual gauge
transformations, under which the {\it connection}  1-form field
$A$
transforms as $$\d A=d_A\omega, \eqno(3.3)$$ while the
$\G$-valued
$(n-2)$-form field $B$ transforms ``vectorially" $$ \d
B=[B,\omega],
\eqno(3.4)$$ where $\omega$ is the gauge parameter, and hence is
a
${\cal G}$-valued 0-form field.  The differential operator $d_A$
is
the exterior
 covariant derivative operator with respect to the connection
$A$:
$$ d_A\omega:=d\omega+[A,\omega], \eqno(3.5)$$ and $[\,,\,]$ is
the commutator of the ${\cal G}$-valued forms.
 \par
Besides the ``$\omega$-transformation", BF type topological
field theories admit  as a symmetry another transformation,
namely
$$\delta A=0, \eqno(3.5)$$ $$\delta B=d_A \tau, \eqno(3.6)$$
where
$\tau$ is a $\G$-valued $(n-3)$-form field.  \eject
\section{3.1.}{The Parametrization} \par Consider any fixed
connection $\tA$ with values in a lie algebra $\cal G$, such that
$F[\tA]=0$; i.e. $\tA$ is flat. It is possible to do a
generalized
Hodge decomposition[18] for any $p$- form $B_{(p)}$ that takes
its
values in the fibres of a flat vector bundle with connection
$\tA$.
To accomplish this, first define $d_{\tA}$, the covariant
exterior
derivative with respect to the flat connection $\tA$, by:  $$
d_{\tA}
B_{(p)}:=d B_{(p)} + [\tA, B_{(p)}], \eqno(3.7)$$ acting on a
$p$-forms $B_{(p)}$. The corresponding covariant co-exact
derivative
is:  $$ \dltA:=(-1)^{np+n+1} *\dA*  \,\, .  \eqno(3.8)$$ Since
$\tA$
is flat, both $\dA$ and $\dltA$ are nilpotent.  The generalized
Hodge
decomposition of $B_{(p)}$ is:  $$ B=B^H
+d_\tA\beta^T_{(n-3)}+\d_\tA\gamma^L, \eqno(3.9)$$ where the
harmonic
form $B^H$ satisfies $$\d_\tA B^H=0=d_\tA B^H, \eqno(3.10)$$ the
transverse form field $\beta^T_{(n-3)}$ satisfies:
$$\delta_{\tA}\beta^T_{(n-3)}=0.  \eqno(3.11)$$ Finally, the
longitudinal field $\gamma^L$ satisfies $$d_\tA\gamma^L=0.
\eqno(3.12)$$
 \par Given an arbitrary connection 1-form field $A$ (which in
general is not flat), parametrize the connection 1-form field $A$
as
follows:\footnote{*}{Elitzur et.  al.[21] arrive at a
similar expression for flat connections on two dimensional
manifolds.}:  $$A={\tA} + \delta_{\tA} \a^L, \eqno(3.13)$$ where
$\a^L$ is a longitudinal 2-form, and $$\tA :=U^{-1}\th
U+U^{-1}dU,
\eqno(3.14)$$ now refers to the flat part of the connection $A$,
so
that $$F(\tA):=d\tA+\half[\tA,\tA]=0.$$ Further, $\th$ is a gauge
invariant
 1-form field such that $$  d\th+[\th,\th] =0, \eqno(3.15)  $$
and
 $U$ is a $G$-valued function over $M$. In terms of this
decomposition, the curvature of $A$ is:  $$F(A)=d_\tA \delta_\tA
\a+\half[\delta_\tA\a,\delta_\tA\a].  \eqno(3.16)$$ It is
important
to note the following:  \item{i.} The decomposition (3.13) is
unique
if and only if $F(A)=0$ implies that $\alpha^L=0$. We will
henceforth
assume this to be the case, and discuss the implications of
non-trivial solutions in the concluding section; \item{ii.} The
condition (3.15) follows from the flatness of $\tA$.  \par In the
next section, we will use this parameterization for the $A$ and
$B$
fields in order to compute the partition function for non-abelian
BF
theories in $n$ dimensions.
\section{3.2.} {The Partition
Function
for Non-Abelian BF Theories} \par  The partition
function for the quantum field theory is defined by:  $$Z=\int
{\dmu[A]\dmu[B] \over V_G} \exp{iS[A,B]},\eqno(3.17)$$ where in
general $\dmu[\omega_{(p)}]$ denotes the functional measure on
the space
of Lie-algebra valued $p$-forms that normalizes the Gaussian
functional integral:  $$ \int \dmu[\omega_{(p)}] \exp
i<\omega_{(p)}\vert
\omega_{(p)}> =1 \,\,\,,$$
with inner product now defined by $$
<\omega_{(p)}\vert\omega_{(p)}>:= \hbox{Tr}\int_M \omega_{(p)}
\wedge * \omega_{(p)}
\,\,\, .  \eqno(3.18)$$ The trace above is with respect to the
adjoint reppresentation of $G$. \par We begin with the
decomposition
introduced in the first section.  We first change variables from
$(A,B)\rightarrow(\tA,\alpha^L,
B^H,\gamma^L_{(n-1)},\beta^T_{(n-3)})$ The resulting form of the
partition function is:  $$ Z= \int \dmu[\tA] \dmu[B^h]
\dmu[\beta^T]
\displaystyle{1\over V_G}
   J[\tA] Z'[\tA] \,\,\, , \eqno(3.19)$$ where $J[\tA]$ is the
Jacobian of the transformation to the generalized Hodge
components:
$$\eqalignno{ J[\tA]&:= \fd^\half (\dA\dltA)_{(2)}
\fd^\half(\dltA\dA)_{(n-3)} \fd^\half(\dA\dltA)_{(n-1)}\cr
& =
\fd (\dA\dltA)_{(2)} \fd^\half(\dltA\dA)_{(n-3)} \,\, .
&(3.20)\cr}$$ The last line in the above is obtained by using the
identity (A.12). In Eq.(3.19) we have defined the functional
integral:  $$ Z'[\tA]:= \int \dmu[\alpha^L]\dmu[\gamma^L] \exp i
<B^H+\dA\beta^T + \dltA\gamma^L\vert \dA\dltA\alpha^L+
[\dltA\alpha^L,\dltA\alpha^L] > \,\,\, .  \eqno(3.21)$$ Although
$Z'$
appears to depend on $B_H$ and $\beta^T$ as well as $\tA$, we
will
now show that it depends only on $\tA$, hence justifying the
notation.  \par
The
integration over $\alpha^L$ and $\gamma^L$ in (3.21) can be done
as
follows. First expand $\alpha^L$ in a complete, orthonormal basis
of
Lie-algebra-valued 2-form eigenmodes of the operator
$(\dA\dltA)_{(2)}$:  $$ \alpha^L = \sum_i C_i \ps^i,\eqno(3.22)
$$
where $$ (\dA\dltA)_{(2)} \ps^i = \Lambda^i \ps^i,\eqno(3.23)$$
and
$$ <\ps^i\vert \ps^j> =\delta^{ij}.\eqno(3.24)$$ Similarly,
expand:
$$ \gamma^L_{(n-1)}=\sum_j f_i \ph^i,\eqno(3.25)$$ where the
$\ph^i$
are normalized eigenmodes of $(\dA\dltA)_{(n- 1)}$, corresponding
to
eigenvalues $\lambda_i$.  \par We now prove that $\ph^i$ can be
chosen so that $$ *\dltA\ph^i =
\sqrt{\Lambda^i}\ps^i.\eqno(3.26)$$
\def\pstilde{\tilde{\psi}_{(2)}} First define
$\pstilde^i:=-*\dltA\ph^i$, and then evaluate $$\eqalignno{
(\dA\dltA)_{(2)}\pstilde^i&=\dA\dltA * \dltA \ph^i\cr
   &=(-1)^{n-1}**\dA**\dltA*\dltA \ph^i = -*\dltA\dA\dltA
\ph^i\cr
   &=-\lambda^i*\dltA\ph^i=\lambda^i\pstilde^i.&(3.27)\cr}$$ This
implies that $\pstilde^i$ are eigenmodes of $(\dA\dltA)_{(2)}$,
with
eigenvalue $\lambda^i$. Similarly, one can show that
$*\dltA\ps^i$
are eigenmodes of $(\dA\dltA)_{(n- 1)}$ with eigenvalues
$\Lambda^i$.  This implies that one can always choose a basis in
which $\ps^i\propto *\dltA\ph$, with the constant of
proportionality
in (3.26) determined by the normalization conditions.  \par In
terms
of the above decompositions, $Z'$ can be written:  $$ Z' = \int
\prod_i dC_i \exp i<B^H+\dA\beta^T\vert*F[A]> \int \prod_j df_j
   \exp if_j X^j(C),\eqno(3.28)$$ where we have defined $$
X^j(C):=
\sqrt{\Lambda_j}\left(C_j\Lambda_j + <\ps^j\vert
  \sum_{k,l}[C_k\ps^k,C_l\ps^l]>\right).\eqno(3.29)$$ The
integration
over the coefficients $f_i$ can now be trivially performed to
give a
delta function: $\delta(X^j(C))$.  We now need to assume that
$X^j(C)$
does not vanish for any values of $C_i$ other than $C_i=0$.  This
is
the same as requiring that the operator $d_A\dltA$ not have any
zero
eigenmodes, which was necessary in the previous section to ensure
the
uniqueness of the decomposition. Although this may not be true
globally, we can restrict the region of integration to the
neighbourhood of $C_j=0$, in which case it is plausible that the
term
quadratic in the $C$'s can never cancel the term linear in $C$'s.
In
this case, the functional integration over $\alpha^L$ has support
only
at $\alpha^L=0$, and, using  (3.23), one obtains:  $$
Z'[\tA]=\fd^{-{3\over2}}(\dA\dltA)_{(2)},\eqno(3.30)$$ which
depends
only on $\tA$ as claimed. This is the crucial result of this
paper,
since it allows the rest of the partition function to be
integrated
in a straightforward fashion.  \par From (3.30), the partition
function becomes $$Z=\int\dmu[\tA]\dmu[B^H]\dmu[\beta^T]{1\over
V_G}\fd^{-
\half}(\dA\dltA)_{(2)}\fd^\half(\dltA\dA)_{(n-3)}.\eqno(3.31)$$
We
now parameterize $\tA$ according to (3.14).  The Jacobian for
this
change of variable can be computed from the inner product $<\d
\tA\vert \d \tA>$, where the variation $\d \tA$ is $$\d
\tA=\dA\d\lambda+U^{-1}\d\th U,\eqno(3.32),$$ with the variation
$\d\lambda:=U^{-1}\d U$.  The flatness of $\tA$ implies that
(3.32)
in fact coincides with the Hodge decomposition of its variation
(which is after all a cotangent vector).  Hence $U^{-1}\d\th U$
must
be the harmonic part of $\d\tA$, with the consequence that
$$\dA\d (U^{-1}\th U)
=0=\dltA\d (U^{-1}\th U).\eqno(3.33)$$ The result is that
$<\d\tA\vert\d\tA>$
has no cross-terms in $\d\lambda$ and $\d\th$ and hence the
Jacobian
for $\tA\rightarrow (U,\th)$ is $\fd^\half(\dltA\dA)_{(0)}$.  The
result for the partition function is
$$Z=\int\dmu[\th]\dmu[B^H]\dmu[U]\dmu[\beta^T]{1\over
V_G}\fd^\half(\dltth d_\th)_{(0)}\fd^{-
\half}(d_\th\dltth)_{(2)}\fd^\half(\dltth d_\th)_{(n-3)}.
\eqno(3.34)$$
where in the above $\dmu[U]$ denotes the Haar measure on the
gauge group, and we have used the gauge invariance of the
functional determinants to
replace the various covariant derivative operators with respect
to
$\tA$ by covariant derivative operators with respect to $\th$.
\par
It is convenient at this stage to define precisely what is meant
by $V_G$, the ``volume of the group of gauge transformations." This
must be defined so as to eliminate the unphysical gauge modes in
the functional integral. By performing the above reparametrizations
we have effectively isolated these modes: they are contained in the
group variables $U$ and the transverse $(n-3)$-forms $\beta^T$. As
in the the abelian case, it would not be consistent with
perturbation theory to simply eliminate the integrals over these
variables. In complete analogy with the discussion in
Section 2, the correct choice for $V_G$ is therefore:
$$V_G=V^{(n-3)}_A V_U,\eqno(3.35)$$ where
$$V_U:=\int \dmu[U],\eqno(3.36)$$ and $$V^{(n-3)}_A:=\int
{\dmu[\beta^{(n-3)}]\over V_A^{(n-4)}}.\eqno(3.37)$$ In the above,
it
is
important to notice that we functionally integrate over all the
$(n-3)$-forms, and not just the transverse modes.  We then divide
by
the volume of the orbits of gauge transformations
 acting on $(n-3)$-forms.  The rest of the calculation, namely
the
integration over $U$ and $\beta^T$, proceeds as in section 2.3
after
equation (2.31).  The reason for this is that a small
neighbourhood
in the hypersurface (in configuration space) of flat connections
has
the same structure as the hypersurface of closed forms in the
configuration space of abelian BF theory.  The result is
therefore
identical in form to (2.37), namely:  $$Z=\int
\dmu[\theta]\dmu[B^H]\prod_{i=2}^{n} \fd^{\half
(-1)^{i+1}}(\dltth d_\th)_{(n-i)}\quad
 \fd^{\half}(\dltth d_\th)_{(0)}.\eqno(3.38)$$ The last factor in
the
above is square root of the functional determinant of the
laplacian
acting on 0-forms with respect to the connection $\th$,and it
occurs
in $Z$ only for {\it odd}-dimensional
 spacetimes.  We can again use identities (A.3) and (A.12) to
simplify (3.38) for odd $n$ to obtain:
$$Z=\int\dmu[\th]\dmu[B^H]\prod_{j=0}^{m}\fd^{(-
1)^j(m-j+\half)}\Delta_j(\th),\eqno(3.39)$$ where $\Delta_j(\th)$
is
the Laplacian $(\dltth d_\th+d_\th\dltth)_j$.  In the case of
{\it even}
$n$, the same identities and (A.2) yield the result $Z=1$, i.e.,
the
triviality of the Ray-Singer torsion in even dimensional
manifolds.
\par
 Note that the integrand in (3.39) is in general a function of
the
$\th$, the coordinates in the space of equivalence classes (under
gauge transformations) of flat connections on $M$.  Hence, the
integration is over the bundle ${\cal N}$ to the {\it moduli
space}
${\cal M}$ over $M$, with fibers isomorphic to the cohomology
space
$H^{(n-2)}(M)$.  In particular, consider the case $n=3$, with the
gauge group $G$ the Lorentz group SO(2,1), so that the BF theory
is
general relativity in 2+1 dimensions.  In this case the space
${\cal
N}$ is just the cotangent bundle to ${\cal M}$[15].  \par For
$n=3$,
our result corresponds to the expression obtained by Witten in
Reference 15: One simply uses the result (A.11) to obtain
$$Z=\int\dmu[\th]\dmu[B^H]\fd^{-\half}\Delta_{(1)}(\th)\quad
\fd^{3\over2}\Delta_{(0)}(\th),\eqno(3.40)$$ and then notes that
the
functional determinant of Witten's operator $L_-$ can be written
as
$$\fd(L_- )=\fd^\half\Delta_{(1)}(\th)\fd^\half\Delta_{(0)}(\th).
\eqno(3.41)$$ \eject \section{4.}{Conclusions} \par We have
expressed
the partition function for non-abelian and abelian BF type
topological field theories (and for abelian Chern-Simons theory)
in
terms of functional determinants of the appropriate differential
operators.  This was achieved by introducing appropriate
parameterizations of the fields, suggested by the geometry of the
configuration spaces in question. Our procedure avoided many of
the
complications (in particular the introduction of non-physical
ghost
fields and ghosts for ghosts) associated with the technology of
gauge-fixing and provided a simple expression for the volume of
the
gauge orbits for both abelian and non-Abelian B-F theories (cf.
Eq.(2.36)).  \par In the Abelian case, our parameterizations
consist
of precisely the Hodge decompositions for the fields, as
suggested by
Horowitz and Srednicki[12] and by Blau and Thompson[2], and our
results agree with those obtained in the latter paper and by
Schwarz in [13], though we
presented them in a somewhat simpler form.  In addition, we were
able
to derive the identities necessary to prove the triviality of the
partition function directly from the properties of the Hodge
decomposition, without using functional integrals, as was done in
Ref.2.  \par For non-abelian theories, the (Lie algebra-valued)
vector field $B$ was parameterized by its Hodge decomposition
with
respect to the flat connection $\tA$.  Our parameterization of
the
connection 1-form field $A$ is rather more speculative, in the
sense
that we must remain in a neighbourhood of the flat connection
$\tA$
sufficiently small so that the solutions of $F(A)=0$ are all at
most
gauge transformations of $\tA$.  In this case, we were able to
explicitly compute the partition function for any dimension of
the
spacetime manifold in terms of functional determinants of twisted
Laplacians with respect to the flat connections $\tA$.  In other
words, the partition function (3.39) is an integral over some
bundle
over
the moduli space of flat connections of an integrand which is
essentially
the Ray-Singer torsion with respect to that flat
connection.\footnote{$^1$}{It is perhaps worth noting that this
result can be reformulated as a statement that the semi-classical
approximation in such theories is exact.}  We
showed that, as expected, the partition function is trivial in
the
case of even dimensional spacetimes.  For three dimensional
spacetime
our partition function agrees with that obtained by Witten[15].
As
far as we know ours is the first explicit result for dimensions
greater than four.  \par We are currently working on the
extension of
our method to calculate partition functions for super BF theories
and
other topological field theories.  Furthermore, we are attempting
to
explicate the structure of configuration space in more detail in
order
to improve the applicability of our parameterization.  \eject
\noindent{\bf
ACKNOWLEDGEMENTS} \par We wish to thank S. Carlip, B. Dolan, P.
Kelly
and D.J. Toms for useful suggestions and discussion.  J.G. is
grateful to the Winnipeg Institute of Theoretical Physics for its
kind hospitality and support during the latter stages of this
work.
We also wish to acknowledge the partial support of the Natural
Sciences and Engineering Research Council of Canada.
\eject\noindent{\bf APPENDIX}
\par
We now prove three important identities
used in the text.  In this section the differential operators $d$
and
$\d$ refer to the exterior covariant derivative with respect to
any
flat connection, $\tA$, and its co-derivative, respectively,
acting
on lie-algebra valued forms. The results are of course equally
valid
for the Abelian counterparts of these operators used in Section
2.
For simplicity, however, we have suppressed the subscript $\tA$.
We
shall prove:
$$ \fd[*\d d\d] = \fd^\half[d\d d **\d d\d] =
\fd^{{3\over2}}[d\d]; \eqno(A.1) $$
$$ \fd(d\d)_{(p)}=\fd(\d
d)_{(p-1)}; \eqno(A.2) $$ $$ \fd\Delta_{(p)}=\fd(\d d)_{(p)} \fd
(\d
d)_{(p-1)}.  \eqno(A.3) $$
Note that on the right hand side of (A.1), we have dropped a
field independent phase.
\par
First consider an invertible operator
$O$ that maps the space of
$p$-forms onto the space of $p'$-forms. Given an inner product on
both the space of $p$ and $p'$ forms, the adjoint $O^{\dag}$ of
$O$ is defined by:
$$ <O\omega_{(p)}\vert
\omega_{(p)}'> = <\omega_{(p)}\vert O^{\dag}
\omega_{(p)}'>.\eqno(A.4) $$
In general the determinant of such an operator is ambiguous, even
if it is invertible, because one needs to define separate bases
for the space of $p$-forms and the space of $p'$-forms, so that
invariance is lost\footnote{$^1$}{We are grateful to Steve Carlip
for bringing this fact to our attention}. In the present case,
however, the determinant in question is essentially {\it defined}
by the path integral that we need to evaluate, namely (cf.
Eq.(2.26)):
$$
\fd^{-1}(O)\equiv
  \int\dmu[\omega_{(p)}]\dmu[\omega_{(p')}]
   \exp i<O \omega_{(p)}\vert \omega_{(p')}> \,\,\, .
\eqno(A.5)
$$
\par
Since $O$ is assumed invertible, we can parametrize the space of
$p'$-forms in terms of $p$-forms:
$$
\omega_{(p')}= O \omega'_{(p)}\,\, .
$$
The resulting Jacobian, J(O), can be evaluated by demanding that:
$$\int\dmu[\omega'_{(p)}]\exp i <O\omega'_{(p)}\vert
     O\omega'_{(p)}>
=\int\dmu[\omega_{(p')}] J^{-1}(O)
      <\omega_{(p')}\vert\omega_{(p')}>\,\, .
$$
After performing the two Gaussians in the above expression, this
yields:
$$J(O)= \fd^\half (O^\dagger O)\,\, .
\eqno(A.6)$$
Performing the change of variables, and inserting this Jacobian
in  the functional integral (A.5) gives:
$$\eqalignno{
\fd^{-1}(O)& =
  \int\dmu[\omega_{(p)}]\dmu[\omega'_{(p)}]
   \fd^\half (O^\dagger O)
   \exp i < O \omega_{(p)}\vert O\omega'_{(p)} >\cr
   &=\int\dmu[\omega_{(p)}]\dmu[\omega'_{(p)}]
   \fd^\half (O^\dagger O)
   \exp i <\omega_{(p)}\vert O^\dagger O\omega'_{(p)} >\cr
  &=\fd^{-\half}(O^\dagger O)\,\, .
  &(A.7)\cr}$$
(A.7) is of course  consistent with the definition of $\fd (O)$,
used in previous work[2,12,13].
\par
If we now consider the operator $O=*\d d\d$, then it is easy to
verify that the adjoint is $O^{\dag}=d\d d*$, and (A.1) follows.
\par
To prove the identity (A.2), note that if $\omega_{(p)}$ is an
eigenmode of $(d\d)_{(p)}$ with eigenvalue $\lambda$, then
$\d\omega_{(p)}$, which is a $(p-1)$-form
is an eigenmode of $(\d d)_{(p-1)}$ with the same eigenvalue
$\lambda$.  But all the non- zero eigenmodes of $(\d d)_{(p-1)}$
must
be co-exact, and hence all of its eigenmodes must be of the form
$\d\omega_{(p)}$.  Thus the operators in question in (A.2) have
the
same eigenmodes and eigenvalues, and the result follows.
Finally, to
prove (A.3), we use the orthogonality of the decomposition $$
\Delta_{(p)} = (\d d + d\d)_{(p)}, \eqno(A.8) $$ to prove $$ \fd
\Delta_{(p)} = \fd (\d d)_{(p)}\quad \fd (d \d)_{(p)},\eqno(A.9)
$$
and then use (A.2) to get (A.3).  \par We also note that (A.3)
can be solved for $\fd(\d d)_{(p)}$ and iterated to yield $$
\fd(\d d)_{(p)}
= \prod^p_{i=0} \fd^{(-1)^i} \Delta_{(p-i)}.  \eqno(A.10) $$
Another
useful identity, easily derived from the properties of Hodge
duality
is[12]:  $$\fd\Delta_{(p)}=\fd\Delta_{(n-p)}.\eqno(A.11)$$ Using
this and
(A.2) it is trivial to show that
$$\fd(d\d)_{(p)}=\fd(d\d)_{(n-p+1)}
.\eqno(A.12)$$
\eject \noindent{\bf REFERENCES}
 \item{1.}E.Witten, Comm.Math.Phys. {\bf117}, 353 (1988).
\item{2.}M.Blau and G.Thompson, Ann. of Phys. {\bf 205},130
(1991).
\item{3.} A.Ashtekar, {\it Lectures on Non-Perturbative Quantum
Gravity}, World Scientific, 1991.  \item{4.} E.Witten, Nuc. Phys.
B
{\bf 311}, 46 (1988).  \item{5.} V. Moncrief, J. Math.  Phys.
{\bf
30},2907 (1989).  \item{6.} A. Hosoya and K. Nakao, Class.
Quant.
Grav. {\bf 7}, 163 (1990).  \item{7.} S. Deser, R.  Jackiw and G.
t'Hooft, Ann. of Phys.  {\bf 152}, 220 (1984).  \item{8.} A.
Ashtekar, Class. Quant. Grav. {\bf 6}, L185 (1989).  \item{9.}J.
Nelson and T. Regge, Nuc. Phys. B {\bf 328}, (1989);   S.P.
Martin, Nuc.
Phys. B {\bf 327}, 178 (1990);  S.  Carlip, Phys. Rev. D
{\bf 42}, 2647 (1990).  \item{10.} S. Carlip, Class. Quant. Grav.
{\bf 8}, 5 (1990).\item{11.} G.T. Horowitz, Comm.  Math. Phys.
{\bf
125}, 417 (1989).  \item{12.} G.T. Horowitz and M.  Srednicki,
Comm.
Math. Phys. {\bf 130}, 83 (1990).  \item{13.} A.S.  Schwarz,
Lett.
Math. Phys. {\bf 2}, 247 (1978); Comm. Math. Phys.  {\bf 67}, 1
(1979).  \item{14.} E. Witten, Comm. Math. Phys. {\bf 121}, 351
(1989).  \item{15.} E. Witten, Nuc.  Phys. B {\bf 323}, 113
(1989).
\item{16.}  See for example L.D.  Faddeev and A.A.  Slavnov, {\it
Gauge Fields: Introduction to Quantum Theory}, Benjamin-Cumming,
1980; or M. Daniel and C. M. Viallet, Rev.  Mod. Phys. {\bf 52},
175
(1980).  \item{17.} Y. Choquet-Bruhat, C. DeWitt-Morette and M.
Dillard- Bleick, {\it Analysis, Manifolds and Physics}, Revised
Edition, North-Holland (1982).  \item{18.}  T.  Eguchi, P.B.
Gilkey
and A.J. Hanson, Phys. Reports {\bf 66}, 213 (1980).  \item{19.}
P.K.  Townsend, Phys. Lett. B {\bf 88}, 97 (1979); W.  Siegel,
Phys.
Lett.  B {\bf 93}, 170 (1980).
\item{20} Yu. N. Obukhov, Phys. Letts. {\bf 109B}, 195 (1982).
\item{21.} S.  Elitzur et. al.,
Nuc.
Phys.  B {\bf 326}, 108 (1989).  \end